\documentstyle[epsfig,psfig]{texas}

\begin{document}

\title{Aspects of Inflationary Reconstruction}
\author{Andrew R.~Liddle}
\address{Astrophysics Group, The Blackett Laboratory,\\
Imperial College, London SW7 2BZ, Great Britain}

\begin{abstract}
I review various aspects of techniques for reconstructing the potential of the 
inflaton field from observations, with special emphasis on difficulties which 
might arise. While my view is that if inflation is to prove viable then most 
likely 
it will be one of the simplest models, it is important to consider the impact 
should we need to move to a more complicated model-building realm.
\end{abstract}

\section{Introduction}

One of the most exciting aspects of the rapidly improving observational 
situation in cosmology is the hope that we might learn of processes happening in 
the very early Universe, and thus learn of physics at energies inaccessible to 
terrestrial experiments. A key idea in early Universe cosmology is inflation 
\cite{infrefs}, a 
period of accelerated expansion thought to be driven by the potential energy of 
one or more scalar fields. Assuming all goes well with upcoming experiments, 
particularly satellite projects {\sc MAP} and P{\sc LANCK} aiming to accurately 
measure microwave background anisotropies, one can hope to receive a limited, 
but non-trivial, amount of information concerning the inflationary mechanism. 
That is to say, one can hope to reconstruct part of the inflationary potential 
\cite{CKLL,LLKCBA}. 

Although inflation (or other early Universe ideas such as cosmic strings) is 
often discussed more or less independently of cosmological parameters such as 
the Hubble parameter $h$ and the density parameter $\Omega_0$, it is in fact 
crucial to the entire enterprise of using such observations to constrain 
cosmology. The reason is that, contrary to the impression given by a fair 
fraction of the literature, measurements of the microwave background {\em in 
isolation} tell you nothing about cosmological parameters. This is because 
the influence of the parameters is on the dynamics, whereas the microwave 
background 
anisotropies give us a single snapshot. In order to predict the parameters, we 
need a theoretical prejudice as to the initial conditions, which are processed 
by dynamical evolution into the anisotropies we see. Consequently, fitting for 
the cosmological parameters and for the initial conditions for structure 
formation are not independent tasks which can be decoupled. Rather, they must be 
done together.

\section{The standard paradigm}

In this article I will be discussing a few of the ways in which inflation, while 
being essentially correct, might turn out to be more complicated than envisaged. 
However, I stress that it is probably much more likely, if inflation proves 
correct at all, that it is one of the simpler models which is true. If so, then 
as Neil Turok said at this meeting `The person who fits the data with the fewest 
parameters is the winner' and the game is over. So let's begin by quickly 
reviewing the simplest scenario.

It arises when the dynamics of inflation (both classical and quantum) are 
dominated by a single scalar field evolving in a nearly flat potential. If so it 
is well established \cite{LL92,LL93} that to a good approximation the two types 
of perturbations, scalar and tensor, will take on a power-law form, with the 
tensors giving a subdominant (and quite conceivably negligible) contribution. 
This is certainly expected to be valid for present data, unless we have a 
`designer' model with a very strong spectral feature present on observable 
scales (as discussed in this session by Lesgourgues).

The two power laws require four parameters for their specification. However, 
there is one consistency relation linking the two spectra which means that the 
tensor spectral index is not independent of the other parameters; 
disappointingly this redundancy is unlikely to be useful as almost certainly the 
tensor spectral index cannot be measured anyway. The remaining three parameters 
can be taken as the overall perturbation amplitude, the spectral index $n$ of 
the scalar perturbations and the relative impact $r$ of tensors as opposed to 
scalars on large-angle microwave background anisotropies. In a given model they 
are readily calculated, for example via the slow-roll expansion \cite{LPB}.

\section{Simplest extension: scale-dependent spectral index}

High-accuracy observational data makes stringent demands on theory, so 
eventually the power-law approximation may prove inadequate. There are some 
theoretical reasons to believe that slow-roll might not be all that good; in 
supergravity models the slow-roll parameter $\eta$, which must be small for 
inflation to proceed, takes the form
\begin{equation}
\eta = 1 + \mbox{`other terms'} \,,
\end{equation}
Even if the other terms manage to partly cancel the 1, it may be unlikely that 
they do so to high accuracy. A particular example of this point in action is the 
running-mass models of inflation introduced by Stewart \cite{S97}, where 
slow-roll is due to an accidental, and temporary, cancellation.

If the slow-roll approximation is only weakly satisfied, then higher-order 
corrections \cite{SL93} to the formulae for the spectral index etc become 
significant and have to be accounted for \cite{CKLL2}. More pertinently, the 
power-law approximation is likely to break down \cite{KosT}, and has to be 
replaced by a more general analysis such as a truncated Taylor expansion of the 
spectrum about some scale. (Some kind of expansion must be done to describe the 
spectra with a finite number of parameters which can be fit from the data.) 

We investigated corrections to power-law behaviour in Ref.~\cite{CGL}. Adding in 
extra parameters will always worsen the uncertainty on {\em all} parameters, but 
we 
found that the likely impact on uncertainties in parameters such as $h$ is 
small, while as a bonus we have given ourselves one or more extra inflationary 
parameters to constrain early Universe physics with. In terms of our being able 
to constrain our models, it appears therefore that a breakdown in slow-roll 
should be regarded as a good thing, and we should hope that if inflation is 
correct it is a model of that type.

\section{Isocurvature models}

A much more disastrous turn of events would be if the best models include 
isocurvature perturbations. Many of the most popular inflation models have more 
than one dynamically important field, and as soon as that happens we have the 
possibility of isocurvature modes. These significantly complicate the 
calculation of the microwave background anisotropies, and a particular 
disadvantage is that these models appear to defy reconstruction, in the sense 
that given a set of observations it would be very hard to decide what sort of 
inflation model gave rise to them. One would have to test candidate models 
against the data on a one-by-one basis. Three regimes are possible:
\begin{itemize}
\item {\bf Pure isocurvature models.}\\
The basic idea here is that the field which eventually becomes the cold dark 
matter already exists during the inflationary era, and acquires perturbations by 
the usual mechanism. The idea has quite a long history \cite{KL}, recently 
revived by Linde \& Mukhanov \cite{LM} and by Peebles \cite{P98}.
\item {\bf Mixed adiabatic and isocurvature.}\\
If both modes are present we need more parameters to describe the initial 
conditions \cite{Setal,Ketal}. Calculationally complex; in particular one 
usually needs to know the whole evolution of the Universe after inflation to 
compute predictions, whereas for adiabatic alone one needs evolve only until the 
modes are well outside the horizon.
\item {\bf Low-level isocurvature.}\\
A small isocurvature contribution might not be directly detectable, yet be 
an extra noise source leading to deterioration of cosmological parameter 
determination.
\end{itemize}
Isocurvature models pose two difficulties. The first is that the power spectra 
from the models must be parametrized so they can be fit from the observational 
data, and in such models it is not clear how many parameters one may need to 
introduce, e.g.~treating them as power-laws may be inadequate. More importantly, 
unlike the case of single-field inflation models there is no direct connection 
between these parameters and the inflation model in the form of a set of 
equations. Even if we have a successful fit of model parameters it may be a 
difficult task to deduce the form of the inflation model giving rise to them, 
particularly if details of post-inflation processing of perturbations need to 
be included.

\section{Open inflation models}

Open inflation is another complicated scenario, either in the original
single-bubble Universe models \cite{Gott,Sasetal,BGT} or the more
recently-devised instanton models \cite{HT}.  These models are readily testable
insofar as the geometry of the Universe is measurable, and encouragingly already
the indications are very much in favour of a flat or nearly flat Universe, both
from the microwave background and measurements of the apparent
magnitude--redshift relation for type Ia supernovae. If they do remain viable, 
they pose a similar set of technical problems to those posed in the isocurvature 
case.

\section{Reconstruction without slow-roll}

I end with a separate topic not closely related to the rest of the article.
With the increasingly widespread use of numerical technology in cosmology, and
bearing in mind the possibility that slow-roll may not work all that well, a new
approach to reconstruction is suggested in the single-field case.  The
traditional approach relies on computing a parametrized form of the perturbation
spectra, which can be input into the {\sc CMBFAST} program \cite{SZ} to give the
microwave anisotropy spectrum.  The drawback is that the analytic computation of
the spectra is only approximate, and ultimately this leads to a biased
estimation of the inflaton potential.  This can be circumvented by instead using
exact computation of the spectra, obtained by numerically solving the mode
equations wavenumber by wavenumber as demonstrated in Ref.~\cite{GL}.  These are
input directly into a modified form of {\sc CMBFAST} to give microwave
anisotropy predictions which are exact up to the assumption of linear
perturbation theory.  By obtaining the anisotropies directly from a parametrized
form of the potential, one can estimate the uncertainties in the parameters
describing the reconstructed potential, and the covariances between the
uncertainties on different parameters, directly.  This will be described in more
detail in a forthcoming publication.

\section{Discussion}

This article stresses that models of the initial perturbation spectra are an
integral part of cosmological parameter estimation, and we rely on our present
understanding being a good one.  It is reasonable to hope, and even expect, that
everything will work out quite simply, but I've outlined a few ways in which
things might be more complicated.  On the plus side, I noted that at least if
things are just a little more complicated, especially in the form of departures
from perfect power-law spectra, that is likely to be seen as a good thing as it
increases the amount of readily accessible information about early Universe
physics.

\section*{Acknowledgments}

Thanks to Ed Copeland, Ian Grivell, Rocky Kolb, Jim Lidsey and David Lyth for 
numerous discussions and collaborations related to this work.

\end{document}